\documentclass[aip, amsmath, reprint, numerical]{revtex4-1}
\usepackage{graphicx}

\begin{document}

\title{Hysteresis of nanocylinders with Dzyaloshinskii-Moriya interaction}
\author{Rebecca Carey}
\affiliation{Faculty of Engineering and the Environment, University of Southampton, Southampton, SO17 1BJ, United Kingdom}
\author{Marijan Beg}
\affiliation{Faculty of Engineering and the Environment, University of Southampton, Southampton, SO17 1BJ, United Kingdom}
\author{Maximilian Albert}
\affiliation{Faculty of Engineering and the Environment, University of Southampton, Southampton, SO17 1BJ, United Kingdom}
\author{Marc-Antonio Bisotti}
\affiliation{Faculty of Engineering and the Environment, University of Southampton, Southampton, SO17 1BJ, United Kingdom}
\author{David Cort\'es-Ortu\~no}
\affiliation{Faculty of Engineering and the Environment, University of Southampton, Southampton, SO17 1BJ, United Kingdom}
\author{Mark Vousden}
\affiliation{Faculty of Engineering and the Environment, University of Southampton, Southampton, SO17 1BJ, United Kingdom}
\author{Weiwei Wang}
\affiliation{Department of Physics, Ningbo University, Ningbo 315211, China}
\author{Ondrej Hovorka}
\affiliation{Faculty of Engineering and the Environment, University of Southampton, Southampton, SO17 1BJ, United Kingdom}
\author{Hans Fangohr}
\email{h.fangohr@soton.ac.uk}
\affiliation{Faculty of Engineering and the Environment, University of Southampton, Southampton, SO17 1BJ, United Kingdom}

\date{\today}

\begin{abstract}
The potential for application of magnetic skyrmions in high density storage devices provides a strong drive to investigate and exploit their stability and manipulability. Through a three-dimensional micromagnetic hysteresis study, we investigate the question of existence of skyrmions in cylindrical nanostructures of variable thickness. We quantify the applied field and thickness dependence of skyrmion states, and show that these states can be accessed through relevant practical hysteresis loop measurement protocols. As skyrmionic states have yet to be observed experimentally in confined helimagnetic geometries, our work opens prospects for developing viable hysteresis process-based methodologies to access and observe skyrmionic states.
  \vspace{5mm}
\end{abstract}

\maketitle

There is a continuous demand for developing magnetic data storage devices with higher recording density and improved reliability and robustness. Recent research demonstrates that magnetic skyrmions show great potential to meet such demands,\cite{kiselev2011chiral, fert2013skyrmions, romming2013writing} which drives vigorous research activity to understand the fundamental aspects of their emergence, stability, and manipulability.

Magnetic skyrmions are topologically stable quasi-particles, which have been found to exist\cite{heinze2011spontaneous} with diameters as small as $1\,$nm. They arise in magnetic systems that lack inversion symmetry in the crystal lattice, which gives rise to the chiral Dzyaloshinskii-Moriya interaction (DMI),\cite{dzyaloshinsky1958thermodynamic, moriya1960anisotropic} such as in bulk helimagnetic materials with a non-centrosymmetric crystal lattice,\cite{dzyaloshinsky1958thermodynamic, moriya1960anisotropic} or at the interface between two dissimilar materials.\cite{fert1980role, crepieux1998dzyaloshinsky} Along with skyrmions, DMI may give rise to different types of magnetic spin configurations, including helical and conical structures, all of which have been observed through theory,\cite{bak1980theory, bogdanov1989thermodynamically, bogdanov1994thermodynamically, rossler2006spontaneous} simulation,\cite{kiselev2011chiral, beg2015ground} and experiment.\cite{uchida2006real, muhlbauer2009skyrmion, yu2010real, yu2011near, heinze2011spontaneous}

Sustaining stable skyrmion states in continuous magnetic films requires the application of a significant external field,\cite{muhlbauer2009skyrmion, yu2010real} which has been seen as a disadvantage for developing applications in information storage. As shown recently, however, zero-field isolated skyrmions can be sustained in magnetic nanostructures with confined geometries,\cite{beg2015ground, rohart2013skyrmion} and created through several standard techniques including spin-polarized current injection.\cite{sampaio2013nucleation, lin2013manipulation} The research so far has focused predominantly on two dimensional confined nanostructures with negligible thickness $-$ the influence of which is not yet understood. Indeed, the recent evidence suggests that modulations of magnetization through the thickness of a nanostructure is an important factor determining the stability of skyrmions.\cite{rybakov2013three, beg2015ground, leonov2015chiral}

To explore such thickness dependent magnetization modulations in confined geometries, in this paper we study the hysteresis behavior of nanocylinder structures through full three dimensional micromagnetic simulations. The external field is varied starting from a well defined saturated state to produce a hysteresis loop, and the magnetization patterns recorded along the loop are analysed and classified into accessible states. Such hysteresis loop measurements are a standard laboratory magnetometry protocol, which highlights the practical aspect of our work. We study hysteresis behavior of nanocylinders of a non-centrosymmetric lattice material FeGe. Such systems of 150 nm diameter and 10 nm thickness were demonstrated earlier to sustain stable quasi-uniform states, isolated skyrmions and target states,\cite{beg2015ground} and here we investigate how the diversity and structure of these states evolves with the increased sample thickness. We demonstrate the role of the magnetostatic (demagnetizing) field, which is often neglected in simulations, in determining the overall hysteresis behavior of nanocylinders and show that it aids the skyrmion stability and extends the applied field and thickness range supporting their existence.

\begin{figure*}[t]
  \centering
  \includegraphics[width = \textwidth]{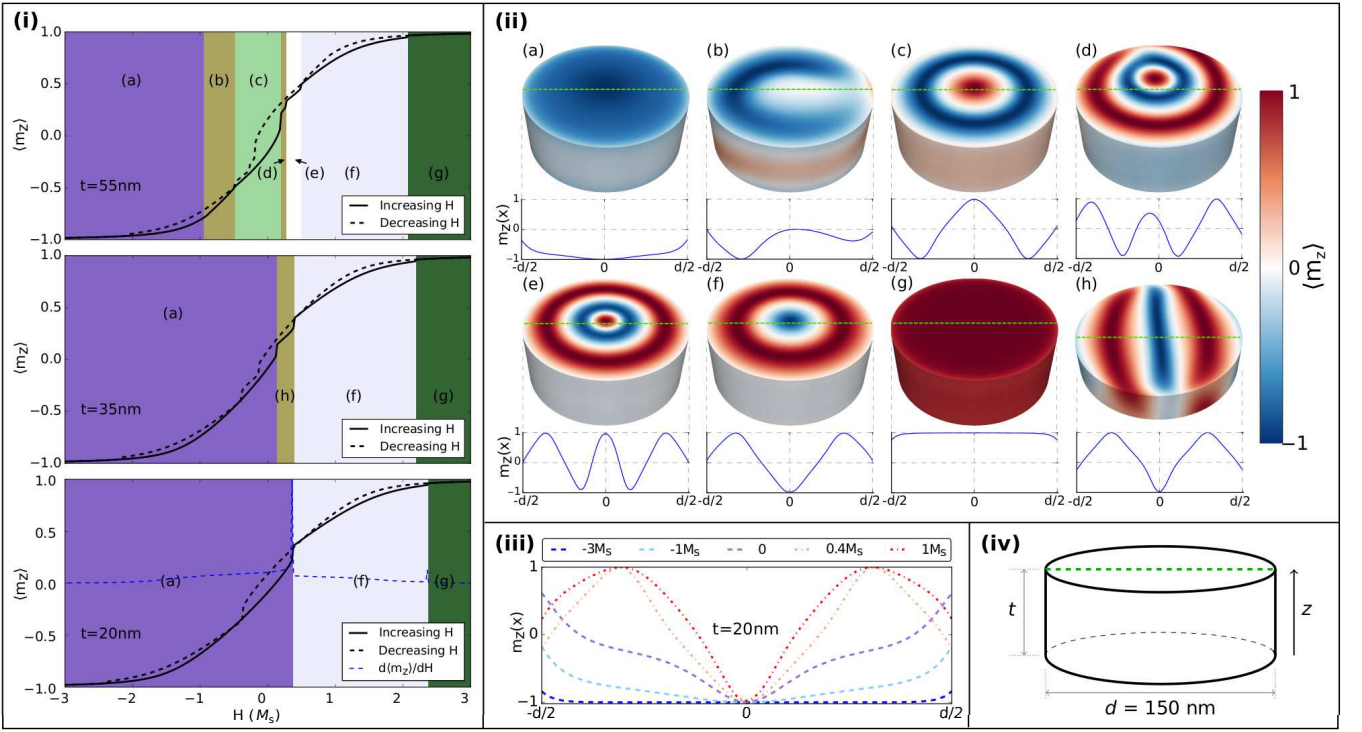}
  \caption{\label{fig:groupedPlots}\textbf{Hysteresis loops and magnetization states of FeGe nanocylinders for different thickness.} \textbf{(i)} Hysteresis curves for nanocylinders with thicknesses $t=20$, $35$ and $55\,$nm. The external magnetic field of strength $H$ was applied in the direction perpendicular to the cylinder base and swept between $\pm 4 M_{\mathrm{s}}$. The different highlighted regions (a)-(h) indicate the range of field values for which different states occurred along the increasing hysteresis loop branch. These states are marked by letters (a)-(h) in \textbf{(ii)}, which shows 3D plots of the thickness-averaged $z$ component of the magnetization, $\langle m_z\rangle$, of typical representative states (a)-(h) and the underlying magnetization profile $m_z(x)$ along the dotted line highlighted across each cylinder. The states (a) and (g) are categorized as incomplete skyrmions, (c) and (f) are isolated skyrmions with cores up and down, (e) is a target state and states (b), (d) and (h), the only states not to exhibit rotational symmetry, are defined as transition states. \textbf{(iii)} Demonstrating how skyrmion and incomplete skyrmion profiles differ at different external field strengths. The figure shows an incomplete skyrmion state (a) at $H=-3M_{\mathrm{s}}$, $-1M_{\mathrm{s}}$, $0M_{\mathrm{s}}$ and a skyrmion state (f) at $H=0.4M_{\mathrm{s}}$, $1M_{\mathrm{s}}$ in a $t=20\,$nm sample. A key feature is decreased tilting at the edges in stronger fields. \textbf{(iv)} Simulated geometry.}
\end{figure*}

The 3D micromagnetic simulations were performed using a finite element approach outlined in \textit{Nmag} \cite{fischbacher2007systematic} and extended to include the DMI term. The simulations integrated the Landau-Lifshitz-Gilbert (LLG) equation using the effective field framework consistent with the following micromagnetic energy:
\begin{equation}
 W[\mathbf{m}] = \int_V [\omega_{\mathrm{ex}}(\mathbf{m})  + \omega_{\mathrm{dmi}}(\mathbf{m}) + \omega_{\mathrm{d}}(\mathbf{m}) + \omega_{\mathrm{z}}(\mathbf{m})] \mathrm{d}V
\end{equation}
where $\mathbf{m} = \mathbf{M} / M_{\mathrm{s}}$ is the normalised magnetization vector, with $M_{\mathrm{s}} = |\mathbf{M}|$ being the saturation magnetization. The micromagnetic energy density includes the symmetric exchange $\omega_{\mathrm{ex}} = A\left[(\nabla m_x)^2 + (\nabla m_y)^2 + (\nabla m_z)^2 \right]$ and DMI energy density $\omega_{\mathrm{dmi}} = D \mathbf{m} \cdot (\nabla \times \mathbf{m})$, with $A$ and $D$ being the exchange stiffness and DMI strength, respectively; magnetostatic energy density $\omega_{\mathrm{d}}$ giving rise to demagnetizing field; and the Zeeman term $\omega_{\mathrm{z}} = -\mu_0M_{\mathrm{s}}\mathbf{H} \cdot \mathbf{m}$, with $\mathbf{H}$ denoting the external field vector.
We considered FeGe nanocylinders with a diameter of $150\,$nm and a variable thickness $t$ (Fig.~\ref{fig:groupedPlots}(iv)). The finite element mesh discretization was set to $3\,$nm and guaranteed to be smaller than any of the relevant micromagnetic length scales. The micromagnetic material parameters of FeGe\cite{beg2015ground} used were $M_{\mathrm{s}}=384 \, \mathrm{kAm^{-1}}$, $A=8.78 \, \mathrm{pJm^{-1}}$ and $D=1.58 \, \mathrm{mJm^{-2}}$. The thickness of the nanocylinders was varied between $10$-$80\,$nm in $5\,$nm increments and a hysteresis loop was computed for each thickness.

The system was first initialised by equilibrating the magnetic state in a negative saturating external field oriented perpendicular to the cylinder base ($z$-axis). The hysteresis behavior was then simulated with a consistent procedure by starting from a well defined negative saturation state, increasing the external field $H$ in fine steps $\Delta H$, equilibrating the system at every field step, and using the previous magnetization state as an initialization in the subsequent field step. In all simulations, the $H$ was applied perpendicular to the cylinder base and swept between the saturating values $H = \pm 4M_\mathrm{s}$. We used $\Delta H=0.02M_{\mathrm{s}}$, after confirming that this step size was sufficient to guarantee reproducible simulations which were not affected by its further reduction.

The examples of hysteresis loops for thicknesses $t = 20$, $35$ and $55\,$nm are shown in Fig.~\ref{fig:groupedPlots}(i) as plots of the spatially averaged $z$-component of magnetization, $\left< m_z \right>$, for different external fields $H$. The highlighted regions (a)-(h) indicate the external field intervals where distinct classes of magnetization patterns were observed during the hysteresis process. Thus overall we found 8 different magnetization configurations (a)-(h), examples of which are shown in Fig.~\ref{fig:groupedPlots}(ii) (a)-(h) and (iii). Below, we classify these configurations into the following four categories.

1. \emph{Isolated skyrmions} (Fig.~\ref{fig:groupedPlots}(ii) (c), (f)). Isolated skyrmions are defined as axially symmetric states that contain one full spin rotation along the diameter in the sample, \emph{i.e.}\ the modulations of the magnetization along the diameter are seen to rotate by at least $2\pi$. Figs.~\ref{fig:groupedPlots}(ii) (c) and (f) show isolated skyrmion states with core up and down, respectively, as typically observed in our simulations along a hysteresis loop in the low and high field regions (c) and (f) (Fig.~\ref{fig:groupedPlots}(i)). Fig.~\ref{fig:groupedPlots}(iii) shows the associated magnetization vs.\ radial position profiles of the isolated skyrmions (f) with core down, at fields $H=0.4M_{\mathrm{s}}$ and $1M_{\mathrm{s}}$ for a cylindrical sample of thickness $t = 20\,$nm. As can be seen, in contrast to theoretical predictions made for skyrmions in infinite systems, the finite size calculations display additional magnetization tilting at the edge, which results from specific boundary conditions consistent with the boundary DMI interaction in isolated geometries.\cite{rohart2013skyrmion} The extent of this tilting depends on the strength of the external field along the hysteresis loop.

2. \emph{Incomplete skyrmions} (Fig.~\ref{fig:groupedPlots}(ii) (a), (g)).
Incomplete skyrmions is a terminology used to refer collectively to axially symmetric states that do not sustain full spin rotation along the diameter. These states have also been termed as quasi-ferromagnetic or edged vortex states.\cite{beg2015ground} Fig.~\ref{fig:groupedPlots}(iii) demonstrates how the magnetisation profiles of these incomplete skyrmions (at fields $H=-3M_{\mathrm{s}}$, $-1M_{\mathrm{s}}$, $0M_{\mathrm{s}}$) differ from isolated skyrmions (at fields $H=0.4M_{\mathrm{s}}$, $1M_{\mathrm{s}}$). The tilting at the edges of the sample in incomplete skyrmions, prevalent even at strong external fields, is again due to the DMI. It is strongly dependent on the external field strength and penetrates into the sample as the field strength decreases.\cite{rohart2013skyrmion, meynell2014surface}

3. \emph{Target states} (Fig.~\ref{fig:groupedPlots}(ii) (e)). Target states are also radially symmetric states and in contrast to isolated skyrmions contain two or more full spin rotations along the diameter, depending on the diameter of the sample and the external field.\cite{leonov2014target} In our examples in Fig.~\ref{fig:groupedPlots}(i) the target states are not seen for samples with low thickness $t = 20$ and $35\,$nm, but are observed at thickness $t = 55\,$nm. As can be seen in Fig.~\ref{fig:groupedPlots}(ii) (e), the target state associated with this thickness sustains two full spin rotations. As will be discussed below, target states appear in our samples when the thickness $t \ge 45\,$nm.

4. \emph{Transition states} (Fig.~\ref{fig:groupedPlots}(ii) (b), (d), (h)). We refer to transition states as the states that do not exhibit rotational symmetry. These include the states which are evident precursors to the isolated skyrmion and target states emerging during the hysteresis loop process (Figs.~\ref{fig:groupedPlots}(ii) (b), (d)), as well as the states resembling the regular `striped' helical state (Fig.~\ref{fig:groupedPlots}(ii) (h)). The latter of these, the helical-like state, does not exhibit the one-dimensional helical structure with uniform propagation direction of regular helices found in infinite geometries and instead, due to the boundary conditions of the confined geometry, partially follows the curvature of the cylinder.\cite{rohart2013skyrmion}

\begin{figure}
  \centering
  \includegraphics[width = 0.5\textwidth]{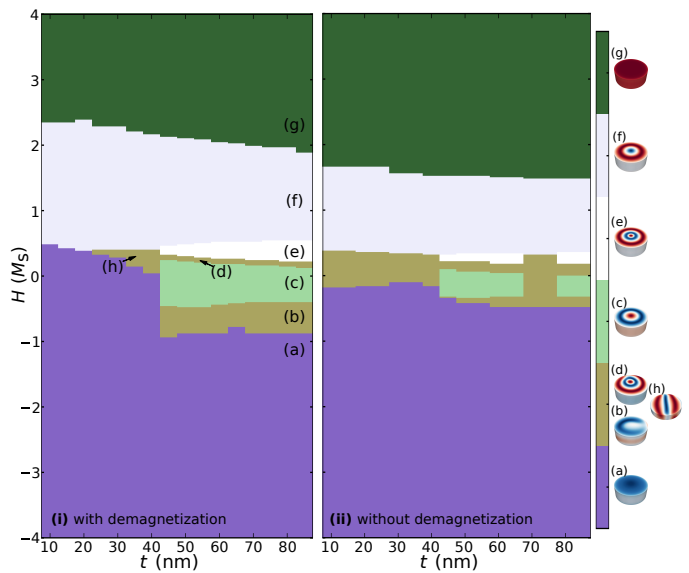}
  \caption{\label{fig:phaseDiagrams}Magnetization state `phase diagrams' showing the type of states sustained in the nanocylinders of different thicknesses $t$ and at different external field strengths $H$ along the increasing branch of hysteresis loop between the saturating fields $H=\pm 4M_{\mathrm{s}}$. (i) full magnetostatics calculations, (ii) no magnetostatics, \emph{i.e.}\ without any demagnetizing field effects.}
\end{figure}

Having defined a terminology for all observed states, we now systematically investigate their occurrence at different sample thicknesses $t$ and external field strengths $H$, in consistency with the major hysteresis loop process run from negative to positive saturation. Such a $t-H$ `state phase' plot is shown in Fig.~\ref{fig:phaseDiagrams}(i) and demonstrates that the magnetization states from the four categories introduced above form well-defined regions.
In the thin geometry range for $t<20\,$nm, the simulated samples support the radially symmetric down and up incomplete skyrmions (a, g) and the isolated skyrmion with core down (f), which are separated by well defined threshold field values. The incomplete skyrmion down state (a) remains into positive fields before switching to the isolated skyrmion (f).
The thickness range $20\,$nm$\,<t<40\,$nm behaves similarly, with the addition of a small field interval with transition states (h) separating the incomplete and isolated skyrmion state phases at smaller and larger fields, respectively.
The behavior becomes enriched at the threshold thickness $t = 45\,$nm, which now allows a sustained isolated skyrmion with core up (c), in addition to the isolated skyrmion with the core down preferred at higher fields, and also the target state (e). The isolated skyrmion with core up (c) is stabilised at negative fields and is seen to exist into the positive field region, and this trend seems to be preserved for higher sample thicknesses. Similarly, the emergence of the target state (e) seems to be preserved when the sample thickness increases. Thus in cylindrical structures with thickness $t > 45\,$nm the `polarity' of the isolated skyrmion can be switched solely via the hysteresis process.
Furthermore, given that the isolated skyrmion and target states are yet to be observed experimentally, our calculations demonstrate that well-controlled hysteresis loop measurements combined with appropriate high resolution magnetic domain imaging techniques might provide a way for the observation of them.
Fig.~\ref{fig:phaseDiagrams}(ii) shows an identical $t-H$ state phase plot to that in Fig.~\ref{fig:phaseDiagrams}(i) but with the magnetostatic field neglected, \emph{i.e.}\ any demagnetizing field effects not present. Overall, the differences between the state phase plots (i) and (ii) in Fig.~\ref{fig:phaseDiagrams} indicate that the demagnetizing field plays a subtle but important role in governing the behavior along a hysteresis loop.
In particular, the transition states are also seen in the low thickness range $t\leq20\,$nm, indicating they become suppressed under the action of demagnetizing fields; the region of fields around $H=0$ supporting the isolated skyrmions with core up (c) becomes narrower; and the ranges of fields supporting the isolated skyrmion with core down (f) and target states (e) are also reduced.
From this we deduce that once in a specific state, the effects of a demagnetizing field act as an augmenting stabilising factor extending the state phase regions associated with the different states. The enriched behavior at thickness $t = 45\,$nm remains present in both Fig.~\ref{fig:phaseDiagrams}(i) and (ii) and so does not seem to be driven magnetostatically and the resulting tendencies towards the flux closure formation.
We note, that the apparent discontinuity in the behavior at thickness around $70\,$nm is due to the similarity of the states, which could not be resolved using our image detection approach based on quantifying the symmetry of magnetization patterns.

\begin{figure}
  \centering
  \includegraphics[width = 0.5\textwidth]{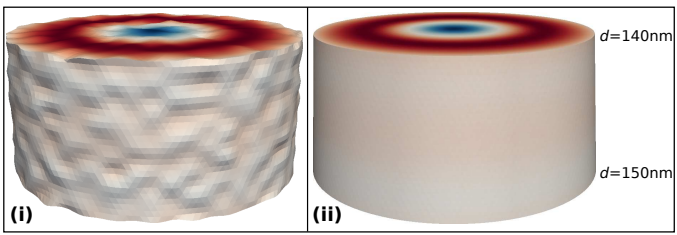}
  \caption{\label{fig:distorted}Demonstrating the distorted geometries used to determine the robustness of the results with respect to irregularities which may arise during fabrication processes, here shown with state (f). The distortion methods were introduced through (i) surface roughness and (ii) tapering the cylinder to form a truncated cone.}
\end{figure}

Additional hysteresis simulations were performed for samples with thicknesses of $40$, $45$ and $70\,$nm, corresponding to the critical regions of the phase diagram, to determine the robustness of the results with respect to irregularities which may arise during fabrication processes.
Specifically, we (1) introduced surface roughness, demonstrated in Fig.~\ref{fig:distorted}(i), by generating randomly distributed surface regions with maximum waviness depth equal to 1\% of the original cylinder diameter, and (2) tapered the cylinder to form a truncated cone geometry (Fig.~\ref{fig:distorted}(ii)) with a maximum base-to-top radius difference of 7\%. The tapering is typical of the etching fabrication process.
We found that the impact of these geometry distortions, in comparison the previous phase plots corresponding to the perfect cylindrical geometry, was a shift to the threshold field values by a maximum of $\pm0.06M_{\mathrm{s}}$, which can be considered negligible to the external field scales of the hysteresis loop.

In conclusion, we have investigated the magnetization behavior in non-centrosymmetric material nanocylinders through micromagnetic simulations of hysteresis loop processes. For the specific choice of 150 nm diameter nanocylinders, the isolated skyrmions and target states emerge in the specific well-defined field intervals along the hysteresis loop if the sample thickness is greater than 45 nm. As these states have yet to be observed experimentally our calculations demonstrate that well-controlled hysteresis loop measurements achievable through standard laboratory magnetometry protocols, and combined with appropriate high resolution magnetic domain imaging techniques, might provide a practical way for their observation. The $t-H$ state phase diagram shown in Fig.~\ref{fig:phaseDiagrams}(i) gives a guide for targeting the individual states. We also demonstrate the subtle effects of the demagnetizing magnetostatic fields in nanocylindrical samples, which are important for the stabilisation of the isolated skyrmion and target states.

The raw data for the figures in this paper, as well Jupyter notebooks \cite{JupyterWebpage} which reproduce the figures are available online for this paper.\cite{GithubrepoSupplementaryMaterial2016}

We acknowledge financial support from EPSRC's DTC grant EP/G03690X/1.

\bibliographystyle{unsrtnat}
\bibliography{paperarxiv}

\end{document}